\documentclass[aps,pre,twocolumn,groupedaddress,showpacs]{revtex4}
\usepackage{graphicx}
\usepackage{color}

\begin{document}

\title{Fluctuation-induced drift in a gravitationally tilted optical lattice}


\author{M. Zelan}
\email[E-mail: ]{martin.zelan@physics.umu.se}
\author{H. Hagman}
\author{K. Karlsson}
\altaffiliation[Current address: ]{Lule{\aa} University of Technology, EISLAB,
971\,87 Lule{\aa}, Sweden}
\author{C. M. Dion}
\affiliation{Department of Physics, Ume{\aa} University, SE-901\,87 Ume{\aa}, Sweden} 

\author{A. Kastberg} \affiliation{Laboratoire de Physique de la
  Mati\`{e}re Condens\'{e}e, CNRS UMR 6622, Universit\'{e} de Nice-Sophia
  Antipolis, Parc Valrose, 06108 Nice Cedex 2, France}

\date{\today}

\begin{abstract}
  Experimental and theoretical studies are made of Brownian particles
  trapped in a periodic potential, which is very slightly tilted due
  to gravity. In the presence of fluctuations, these will trigger a
  measurable average drift along the direction of the tilt. The
  magnitude of the drift varies with the ratio between the bias force
  and the trapping potential. This can be closely compared to a
  theoretical model system, based on a Fokker-Planck-equation
  formalism. We show that the level of control and measurement
  precision we have in our system, which is based on cold atoms
  trapped in a 3D dissipative optical lattice, makes the experimental
  setup suitable as a testbed for fundamental statistical physics. We
  simulate the system with a very simplified and general classical
  model, as well as with an elaborate semi-classical Monte-Carlo
  simulation. In both cases, we achieve good qualitative agreement
  with experimental data.
\end{abstract}


\maketitle

\section{Introduction}
A very general problem in physics is that of a Brownian particle
moving in a periodic potential; a seminal treatment, using
Fokker-Planck formalism, is given by Risken~\cite{Risken:1989}. Of
particular interest is the `tilted washboard potential', where the
Brownian particle is also subjected to a constant force, which can
actually be used to model a wide variety of physical systems (see,
\emph{e.g.},~\cite{Borromeo:2000p1278,Borromeo:2005p1266} and
references therein).  Recently, there has been a significantly
increased interest in the dynamics of small systems, where
fluctuations and noise play a dominating role and where a classical
thermodynamic equilibrium does not occur. There have been theoretical
discoveries
(\emph{e.g.},~\cite{Jarzynski:1997p934,Evans:1993p747,Evans:1994p976,Crooks:1999p977,Seifert:2005p1008})
providing understanding of fluctuations and non-equilibrium
situations, as well as experimental breakthroughs
(\emph{e.g.},~\cite{Wang:2002p1009,Liphardt:2002p1011}). Closely
related to this are systems where noise, or fluctuations, is the
source for directed drift, so-called Brownian motors (see,
\emph{e.g.},~\cite{Reimann:2002p1072,Hanggi:2009p1083,Renzoni:2009p1096}),
or where the noise opens up a possibility for drift in a biased
system, where this bias would otherwise not have been enough to
overcome potential barriers and/or friction.

In this work, we trap and hold cold atoms for several seconds in a
three-dimensional, dissipative optical
lattice~\cite{Grynberg:2001p1099,Jessen:1996p1110}. The thermal energy of the
atoms is of the order of a tenth of the depth of the periodic
potential, and the tilt of the `washboard potential' in the vertical
direction due to gravity is approximately three orders of magnitude
lower than the potential depth (on the range of the period of the
potential)~\cite{gravity}; thus the potential should support the atoms
from gravity with a very good margin. However, these dissipative
optical lattices put the atoms in a regime where fluctuations play a
dominating role, and where dissipation is also present. These
fluctuations will trigger a discernible drift, even with such a small
bias, and threshold effects will be present if parameters, such as
external force, potential depth, fluctuation amplitude, or damping,
are varied. This makes this system a suitable experimental testbed for
fundamental studies of fluctuation phenomena.  In addition, the setup
used here is very close to the double optical lattice arrangement that
has been used to create a Brownian
motor~\cite{Sjolund:2006p1133,Sjolund:2007p1141,Hagman:2008p1134}.
The present work is therefore also of interest for an understanding of
the role of gravity in that context.

\section{The problem in a Fokker-Planck equation context}
A classical particle in the above predicament, with (vertical)
position coordinate $x$, will follow the Langevin equation
%
\begin{equation}
\ddot{x} = -\frac{1}{m}\frac{\textrm{d}}{\textrm{d}x} V(x) - \gamma \dot{x} + \frac{F}{m} + \xi(t) \, .
\label{langevin}
\end{equation}
%
Here, $m$ is the mass, $\gamma$ is a uniform damping constant, $F$ is
a uniform external force, and $\xi$ is a Langevin stochastic
force~\cite{Risken:1989}. The periodic potential is
%
\begin{equation}
V(x) = V_0 \sin (2\pi x/L) \, ,
\label{washboard}
\end{equation}
%
where $L$ is the spatial period of the potential. 

The characteristics of such a system will be determined by the
relative strengths of the terms in Eq.~(\ref{langevin}), and in
particular the magnitude of the friction is important. The case that
compares most closely with our system is one where the friction is
relatively small, or in another terminology, where the system is not
overdamped. This means that the particle can be either in a `locked
state', where it oscillates around a minimum in one potential well, or
in a `running state', where it travels from well to well; and it will
undergo transitions between these states.

The mobility of an ensemble of particles is defined as $\mu=\langle
\dot{x} \rangle / F$ and, for the frictionless case, the locked and
running states would correspond to $\gamma\mu=0$ and $\gamma\mu=1$
respectively. The solution to this general problem is outlined
in~\cite{Risken:1989}. In the case where the friction is small but
still significant, and the noise term is of the same order as the
potential depth (in appropriately rescaled units), the transition from
locked to running, with increased force (or decreased potential
depth), is much less sharp, and the mobility never quite becomes zero,
as long as there is some noise. For the case where the noise term and
the friction are not spatially dependent, there exist analytical
solutions to this general problem~\cite{Risken:1989}.

\section{A tilted dissipative optical lattice}
Dissipative optical lattices arise from atom interaction with a
periodic light shift potential, created by a number of laser beams,
tuned below and relatively close to an atomic, dipole-allowed
transition~\cite{Grynberg:2001p1099,Jessen:1996p1110}. In our case, the
detuning, $\Delta$, is typically of the order of 10--40 natural
linewidths, $\Gamma$, of the transition in question. The proximity to
a resonance means that incoherent light-scattering will be important
for the dynamics of the atoms. There will be diffusion effects
associated with photon recoils and with instantaneous changes to the
light shift potential, due to optical pumping~\cite{Dalibard:1989p1149};
these `heating' effects correspond to the noise term, $\xi$, in
Eq.~(\ref{langevin}). Moreover, with a proper configuration of the
laser beams, laser cooling will be present
(\emph{c.f.},~\cite{Dalibard:1989p1149,Jessen:1996p1110,Grynberg:2001p1099,cool:nobel98});
corresponding to the damping, $\gamma$.

\subsection{Laser cooling}
The seminal treatment of laser
cooling~\cite{Dalibard:1989p1149,Castin:1991} is really only relevant for
atoms that move around in the lattice, and the approximate approach is
there taken that the friction constant, $\gamma$, is a constant and
that a spatial average can be used. This allows for a reasonably
straightforward treatment based on a Fokker-Planck equation approach
\cite{Castin:1991}. However, in actual experiments with dissipative
optical lattices, this cooling mechanism may be relevant for the
initial damping of the thermal energy, and for the first phases of the
route to equilibrium, but the atoms will quickly loose enough thermal
energy in order to be trapped in the potential wells of the lattice;
and at equilibrium, they indeed typically get localized close to the
bottom of the potentials (\emph{c.f.},~\cite{Gatzke:1997p1151}). The
details of the mechanisms for the continued route to equilibrium, for
an atom localized in a well, are not precisely known. However, with
strong support from experimental and theoretical investigations
(\emph{c.f.},~\cite{Castin:1991,Marksteiner:1996p1152,Raithel:1997p1153,%
  Greenwood:1997p1154,Ellmann:2001p1222,SanchezPalencia:2002p1221,Dion:2005p1121}), we
here assume the following. An atom trapped in a well experiences no
direct damping. However, its probability of acquiring energy from
light scattering, and to get unlocked, is higher the more excited it
is in the well. When it gets unlocked, it will be again be exposed to
laser cooling, it will loose its kinetic energy, and it will be
trapped again in some bound state. As this goes on, there will be a
gradual accumulation towards lower lying and more deeply trapped
states, from which the escape probability is low, and eventually an
equilibrium will be reached. Furthermore, the deeper the potentials
are, the larger the portion of atoms that are trapped, but even for
very shallow potentials the majority of the atoms are trapped. Or,
correspondingly, one atom spends most of its time being trapped,
interrupted by short periods of inter-well
flight~\cite{Marksteiner:1996p1152,Ellmann:2001p1222,Dion:2005p1121,Jonsell:2006p1132},
where it can travel over several wells.

\subsection{The damping term in the current work}

In the current work, the relevance of trapping strongly affects how
the damping is to be treated. We make the working hypothesis that when
an atom is trapped (`locked state'), its motion is undamped. If, and
when, it becomes untrapped (`running state'), an effective friction,
$\gamma_\mathrm{LC}$ (with LC standing for ``laser cooling''), turns
on, which we assume can be reasonably well approximated by the spatial
average used in~\cite{Dalibard:1989p1149}, \emph{i.e.}, an untrapped atom
is subjected to dissipation of its momentum as in the traditional
picture of laser cooling.

The acceleration due to the tiny bias force, $F=mg$ (where $g$ is the
gravitational acceleration), is so small that it will not
significantly affect the velocity of the atoms during a single period
of inter-well flight. The damping, $-\gamma_\mathrm{LC}\dot{x}$, will
occur only due to the velocity the atoms acquire from the light
scattering, \emph{i.e.,} the Langevin force, $\xi$, and a free atom
will shortly be trapped again. Thus, the dynamics of the atom will be
of a `stop-and-go' nature. The effect of gravity will only be a very
small average downward drift of the center-of-mass of the sample,
partly because a trapped atom has a slightly higher probability to
escape downwards than upwards, and partly because an untrapped atom
will travel slightly longer distances when going downhill than
uphill. Thus, $\langle x\rangle$ is assumed to change (downwards)
linearly with time, since any memory of the gravitational acceleration
will be erased when the atom is recaptured (see also
\cite{SanchezPalencia:2002p1221}).

\section{Experiment}

For the vast majority of experiments done with dissipative optical
lattice, the holding time in the optical lattice has been rather
short, \emph{i.e.,} 10--100 ms. This is partially because it is
difficult to achieve longer lifetimes, and more importantly because
when the laser cooling dynamics is studied, longer time scales have
not been believed to be important. The basic idea behind our
experiment is to hold the atoms for much longer times, approaching 10
s, and study how the mean position of the sample evolves. This can be
done by direct imaging of the atoms \emph{in situ}. However, as a more
precise diagnostic, we release the atoms and measure their arrival at
a laser probe located at a distance, $l=5$~cm, below the sample
(`time-of-flight detection'~\cite{Hagman:2008p1134}). We do this for a
range of different potential depths, $V_0$, providing us with data for
the mobility as a function of $F/V_0$. The potential depth is varied
by adjusting the irradiances and the detunings of the optical lattice
laser beams ~\cite{Grynberg:2001p1099}. From the time-of-flight data, we
can also analyze the velocity distribution in more detail, in order to
approximately quantify how much time an atom spends in the locked and
the running states on average~\cite{Jersblad:2004p1224}.

\subsection{Experimental set-up}
The experimental set-up has been described in more detail
elsewhere~\cite{Sjolund:2007p1141,Ellmann:2003p1223,Jersblad:2004p1224,Hagman:2008p1134}. In
short, we trap and cool cesium atoms with standard laser cooling
techniques~\cite{cool:nobel98}. The initial cold cloud typically
consists of about $10^8$ atoms with a temperature of around 5~$\mu$K,
which corresponds to about 25$E_\mathrm{rec}$, where $E_\mathrm{rec} =
p_\mathrm{rec}^2/2m$, is the kinetic energy associated with the recoil
of absorption or emission of a single resonant infrared photon (of
momentum $p_\mathrm{rec}$). The atoms are then transferred to a three
dimensional optical lattice, which is constructed by four laser beams
in a three dimensional generalization of the `lin$\perp$lin
configuration'~\cite{Grynberg:1993p1225,Jessen:1996p1110,Grynberg:2001p1099}. The
optical lattice is detuned by either 30 or 40 natural linewidths,
$\Gamma$ ($\Gamma=2\pi\times5.21$~MHz), from the transition between
the hyperfine structure states $F_\mathrm{g}=4$ and $F_\mathrm{e}=5$,
within the D2-line of Cs~\cite{Steck:web}. This configuration will
give a phase-stable lattice with a face-centered-tetragonal geometry,
and with a vertical symmetry axis. The lifetime of the atomic sample
is limited by collisions with the residual gas in the vacuum chamber
and by diffusion in the optical lattice. The maximum lifetime is of
the order of 10 s, giving us ample time to perform the intended
studies.

\subsection{Experimental results}
\begin{figure}[htp]
\centering
\includegraphics[scale=0.25]{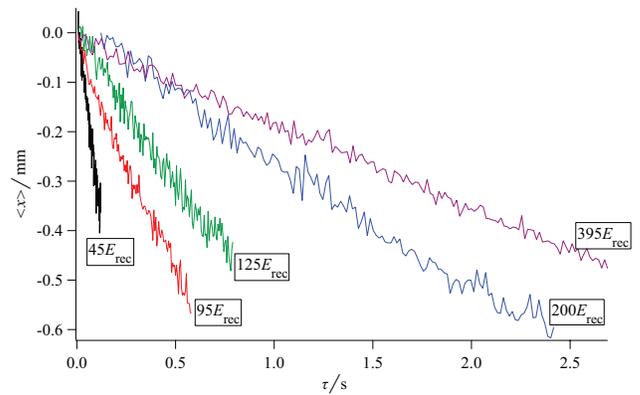}
\caption{(Color online) Position of the center-of-mass of the atomic cloud derived from time-of-flight detection data, as a function of holding time $\tau$ in the optical lattice for the potential depths 45$E_\mathrm{rec}$ (black), 95$E_\mathrm{rec}$ (red), 125$E_\mathrm{rec}$ (green), 200$E_\mathrm{rec}$ (blue) and 395$E_\mathrm{rec}$ (purple), for a detuning of $-40 \Gamma$.}
\label{Fig_1}
\end{figure}

In Fig.~\ref{Fig_1}, we show measured positions of the center of mass
of the atomic cloud as a function of holding time in the optical
lattice, $\tau$, as derived from time-of-flight
data~\cite{Hagman:2008p1134}.
%
%
The arrival time of an atom to the time-of-flight probe is a function
of its initial position and initial velocity. From the average arrival
time, the average position, $\langle x \rangle= \langle \dot{x}
\rangle \tau$, follows directly.  The linear evolution of the drift,
evidencing the `stop-and-go' dynamics, is observed from the data
presented in Fig.~\ref{Fig_1}, where a faster drift due to the
gravitational tilt is also observed for shallower potentials.


\subsubsection{Mobility}
\begin{figure}[htp]
\centering
\includegraphics[scale=0.25]{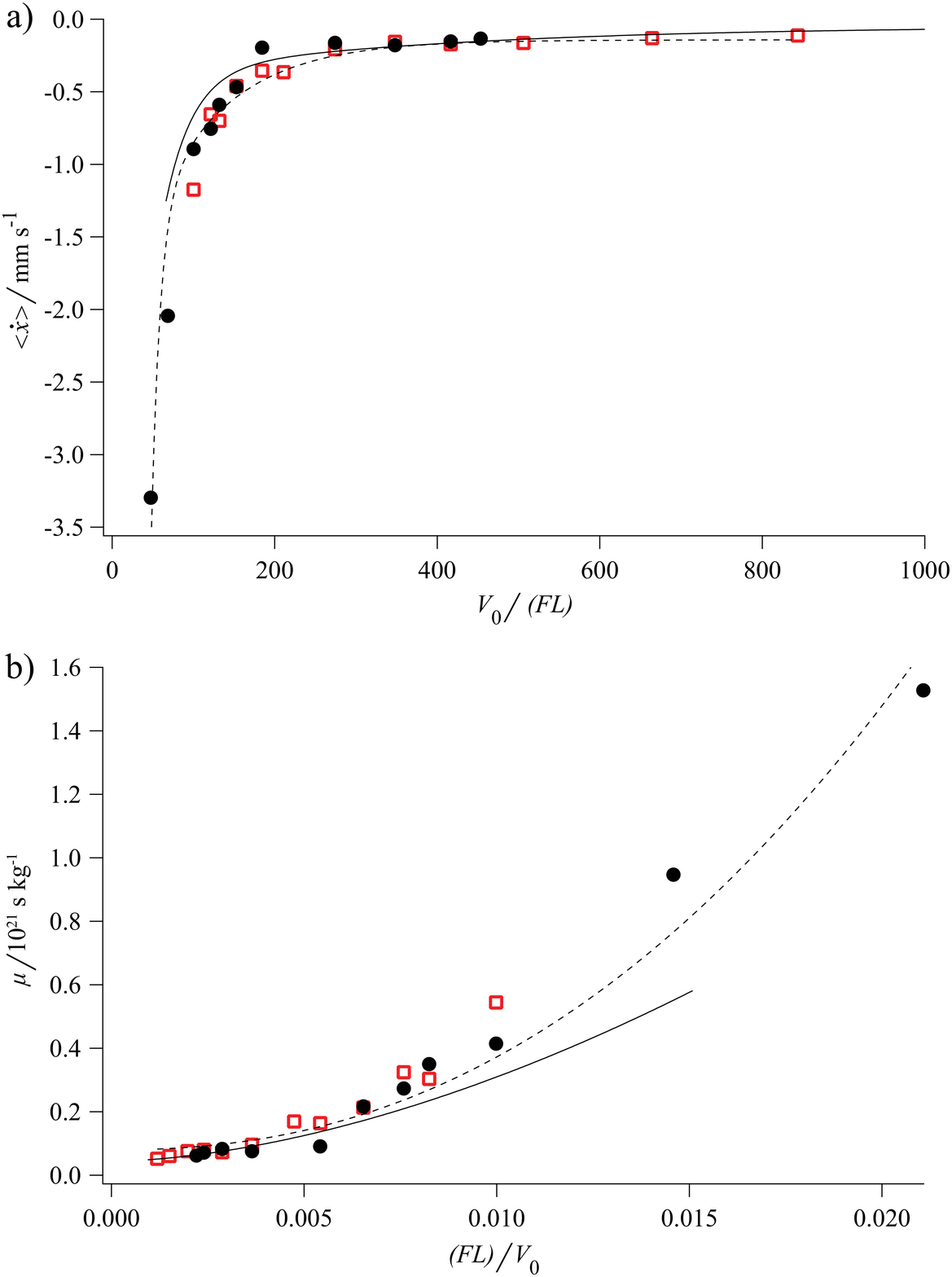}
\caption{(Color online) (a) Derived drift velocity, $\langle \dot{x}
  \rangle$, as a function of potential depth, $V_0$, for the detunings
  $-30 \Gamma$ (open squares), and $-40 \Gamma$ (circles). (b) the
  same data, but plotted as the mobility, $\mu$ as a function of
  $(FL)/V_0$. The solid line is from a semi-classical Monte-Carlo
  simulation, while the dashed line is from a simplified classical
  simulation.}
\label{Fig_2}
\end{figure}

In order to extract the velocity of the drift $\langle \dot{x}
\rangle$, straight lines are fitted to curves like the ones in
Fig.~\ref{Fig_1}. Figure~\ref{Fig_2}a shows results for a range of
data, with $\langle \dot{x} \rangle$ as function of the potential
depth for the detunings $-30 \Gamma$ and $-40 \Gamma$.
%
%
In Fig.~\ref{Fig_2}b we display the same data, but now scaled as the
mobility $\mu$ as a function of the constant force $F$ divided by the
potential depth. This allows for a more direct comparison with the
general theoretical treatment in~\cite{Risken:1989}. Such a comparison
must be made with care, since in~\cite{Risken:1989} a spatially
uniform friction is assumed, whereas our experimental conditions are
such that the damping force that acts on individual atoms depends
strongly on position and kinetic energy. Moreover, even with a
simplified model for a friction force, the coefficient of friction
will vary with detuning~\cite{Dalibard:1989p1149}. Nevertheless,
fluctuation-induced drift in the tilted potential is clearly
demonstrated.


%
%

\subsubsection{Running and locked states}

\begin{figure}[htp]
\centering
\includegraphics[scale=0.25]{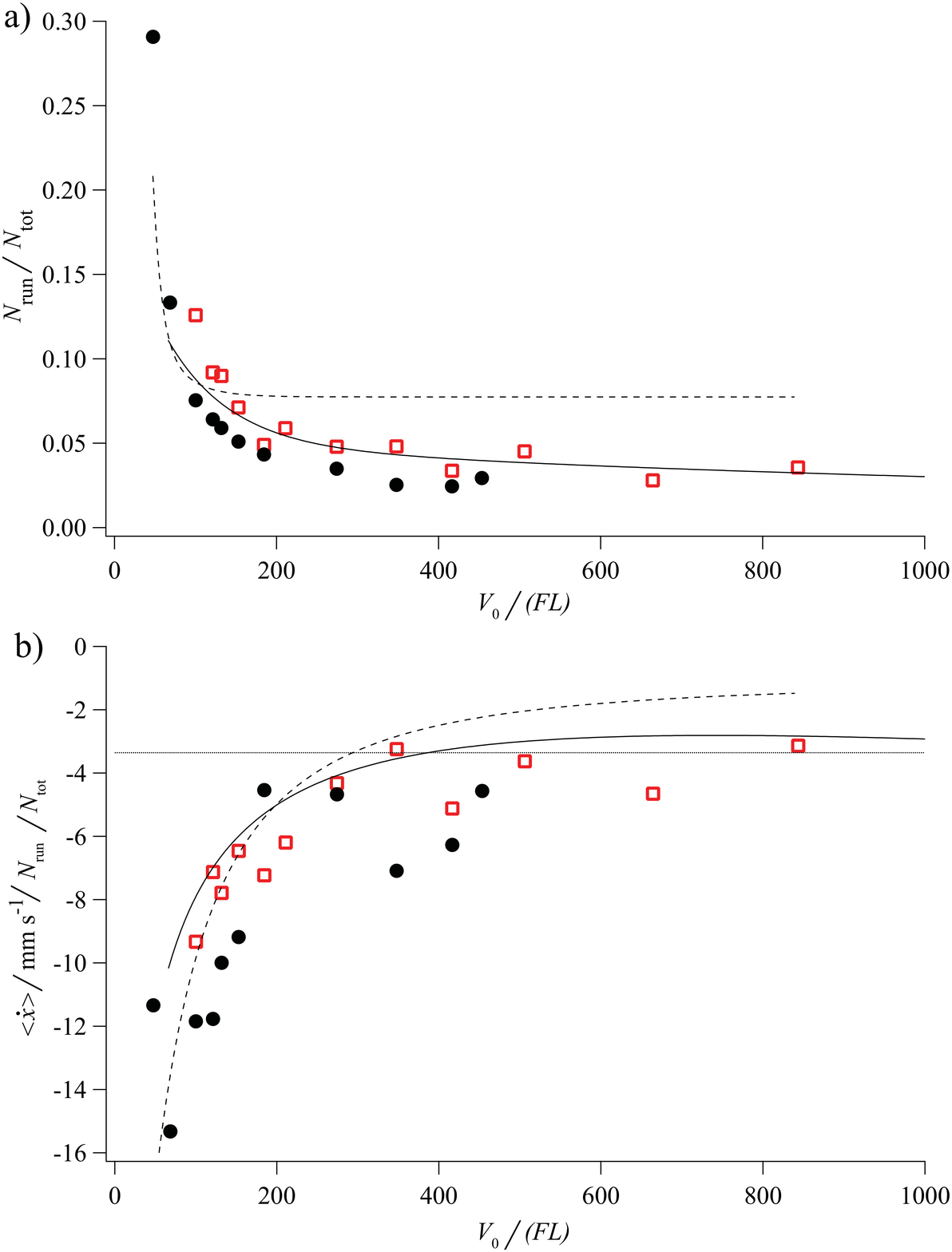}
\caption{(Color online) (a) The fraction of unlocked atoms, as a
  function of potential depth, $V_0$, for the detunings, $-30 \Gamma$
  (open squares), and $-40 \Gamma$ (circles) The solid line represents
  the semi-classical Monte-Carlo simulation, and the dashed line is
  the classical simulation. (b) The drift divided by the fraction of
  atoms is the running state versus the potential depths. A straight
  line indicates the recoil velocity. Note that velocity is defined to
  be positive upwards}
\label{Fig_3}
\end{figure}

By analyzing the velocity distributions obtained by time-of-flight
detection, the fraction of atoms in the running state compared with
the total amount of atoms, $N_\mathrm{run}/N_\mathrm{tot}$ , can be
extracted~\cite{Jersblad:2004p1224,Dion:2005p1121,Hagman:2008p1134}. Assuming that
the momentum distribution corresponds to a (truncated) Gaussian core
of trapped atoms, with wide wings corresponding to untrapped atoms, we
can calculate approximate numbers for $N_\mathrm{run}/N_\mathrm{tot}$
and by Gaussian fits to the momentum distributions. The results of
this for the same data used to observe the drift is shown in
Fig.~\ref{Fig_3}a.
%
%
Within our range of detunings and irradiances, the fraction of the
atoms that are free is typically just a few percent. For potential
depths smaller then about 200$E_\mathrm{rec}$, the fraction of atoms
in the running state increases drastically, and for the shallowest
potentials we use, it gets as high as about 25\%. This behaviour has
striking similarities with Fig.~\ref{Fig_2}a, where the average
velocity downwards due to gravity also increases drastically for the
same potential depths. As an attempt to investigate the dependence
between the drift and the fraction of atoms in the `running state'
further, in Fig.~\ref{Fig_3}b we plot $\langle \dot{x}
\rangle/(N_\mathrm{run}/N_\mathrm{tot})$ versus the potential
depth. From this it is clear that the drift, not being constant,
depends not only on the fraction of untrapped atoms but also on
something else.

\section{Simulations}

\subsection{Semi-Classical Monte-Carlo Method}

With a semi-classical Monte-Carlo simulation of the atom-laser
interaction~\cite{Jonsell:2006p1132,Svensson:2008p1137}, we derive theoretical
data corresponding to all experimental curves shown. In these
simulations, the laser field and the motion of the atoms are treated
classically, which enables tracking of the position and momentum of
each particle, while the internal state of the atom is treated quantum
mechanically, using the true degenerate level structure of the
$F_\mathrm{g}=4 \rightarrow F_\mathrm{e}=5$ transition.  In addition,
the presence of the excited $ F_\mathrm{e}=4$ level is also
included~\cite{Svensson:2008p1137}.  Diffusion and friction arise
``naturally'' from the laser-atom interaction and are position and
velocity dependent, as in the experiment.

The simulations comprise 15000 non-interacting atoms, starting from an
initial cloud contained in a single lattice well, at a temperature of
$5\ \mu\mathrm{K}$.  The drift velocity is obtained by calculating the
median position of the atoms at the end of the simulation and dividing
by the duration of the simulation ($\sim 50\ \mathrm{ms}$).  This is
necessary because a direct calculation of $\left\langle \dot{x}
\right\rangle$ is affected by the presence of a few high-velocity
atoms, corresponding atoms that become untrapped and never get
recaptured by the lattice.  Such atoms are believed to be also present
in the experiment, but drift out of the lattice and never get
detected.

The optical lattice is here one dimensional (along the vertical),
which means that an exact quantitative agreement cannot be expected,
but all qualitative features in the experiment are reproduced, and so
are the orders of magnitude of the mobility, the fractional
populations, and the potential depths where significant features
occur.  Also, the fraction of atoms in the running state is calculated
much more precisely from total energies and positions of the
individual atoms, than it can be determined from the experimental
time-of-flight data.  The results, smoothed out to remove the
fluctuations due to the small sample size, are shown together with the
experimental data in Figs.~\ref{Fig_2} and \ref{Fig_3}.

One of the constraints of the current experimental setup is the use of
gravity as the external force.  This means that the variation of
$FL/V_0$ can only be achieved by a change in $V_0$.  However, $V_0$ is
not independently controllable, but it depends on the irradiance of
the lasers and their detuning with respect to the atomic transition.
These in turn affect friction (Sisyphus cooling) and diffusion (photon
scattering), and therefore also the temperature of the atoms, such
that it can be seen as being a function of the optical lattice
potential depth (see, \emph{e.g.}, \cite{Castin:1991,Grynberg:2001p1099}).

\begin{figure}[htp]
\centering
\includegraphics[scale=0.25]{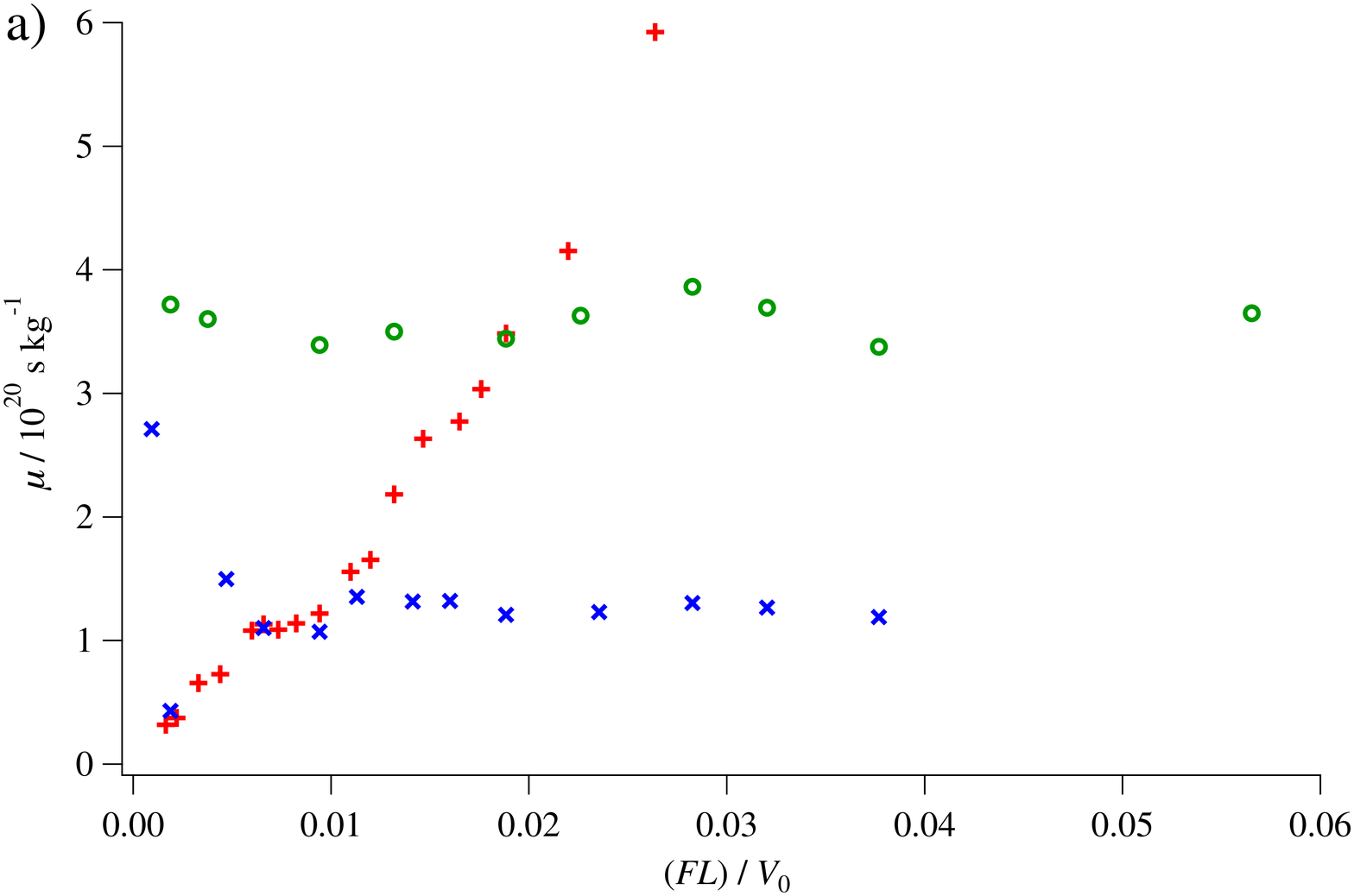}
\includegraphics[scale=0.25]{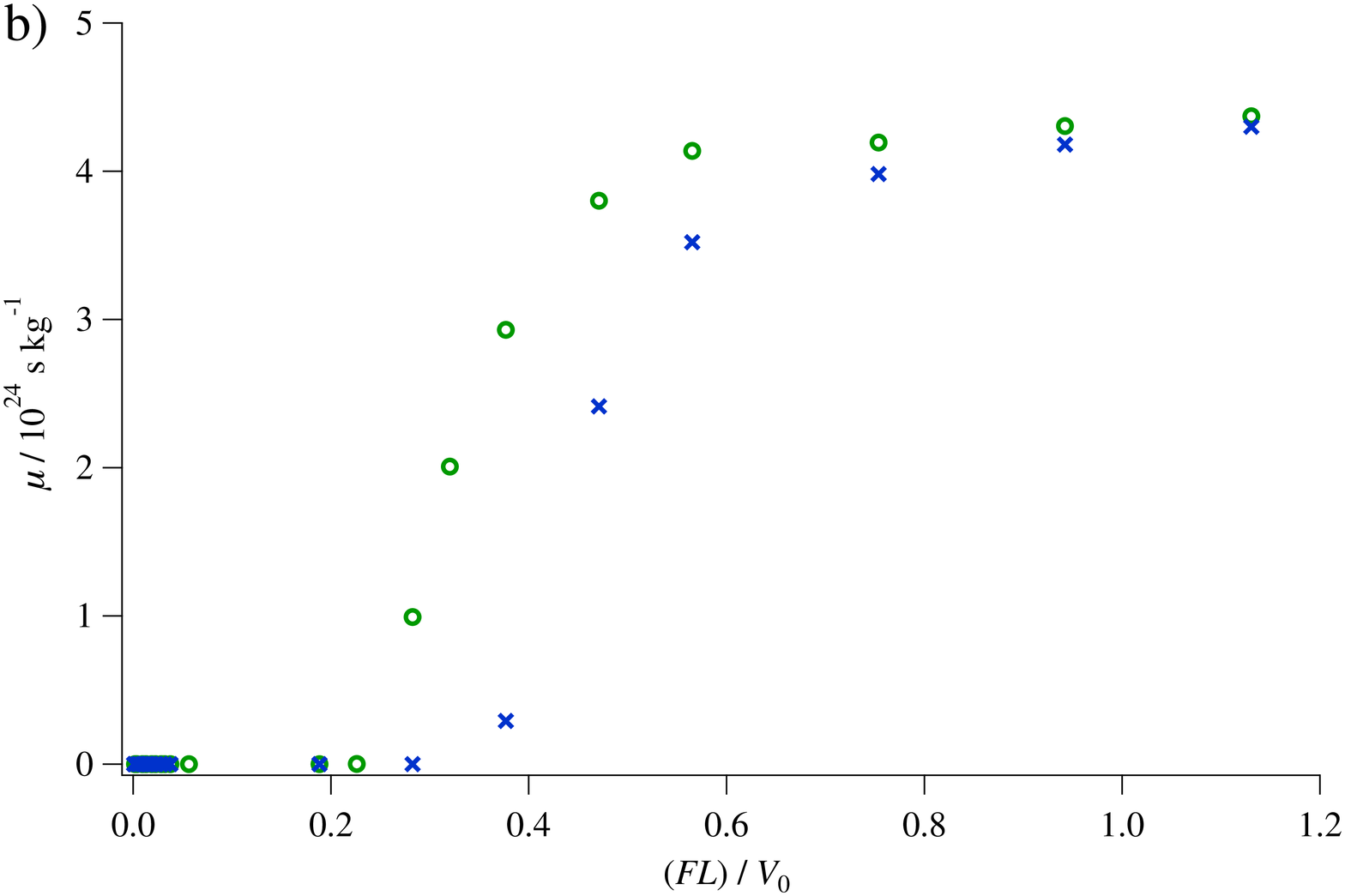}
\caption{(Color online) Mobility calculated from semi-classical
  Monte-Carlo simulations.  $+$: constant force $F = mg$ and varying
  $V_0$ [same data as full line in Fig.~\ref{Fig_2}(b)]; $\circ$:
  varying $F$ and constant $V_0 = 70 E_\mathrm{rec}$; $\times$:
  varying $F$ and constant $V_0 = 140 E_\mathrm{rec}$.  The detuning
  is $-30 \Gamma$.  Panels (a) and (b) differ only by the scaling of
  the abscissa.}
\label{fig:simul}
\end{figure}

To investigate how this affects the mobility, we have run simulations
at constant potential depth for different values of the external
force.  In Fig.~\ref{fig:simul}(a), the simulation result previously
shown in Fig.~\ref{Fig_2}(b) is plotted along with the mobility
obtained varying $F$, for two different values of $V_0$.
Over the range covered by the experiment, the mobility now appears
constant, with a value dependent on the potential depth.  Further
increasing the value of the external force, Fig.~\ref{fig:simul}(b),
the well-known behavior of Brownian particles in tilted potentials
emerges, with a transition region between locked and running
states~\cite{Risken:1989}.  It thus appears that the variation of the
mobility as a function of $V_0$, Fig.~\ref{Fig_2}, reflects the fact
that it is the laser irradiance, and not strictly the potential depth,
that is modified.  In addition, the current experiment probes a region
corresponding to a low-mobility locked state, in accord with the fact
that most atoms have an energy below the well-to-well barrier
height~\cite{Jersblad:2004p1224,Dion:2005p1121}.  Note that the semi-classical
model used here cannot reproduce the Doppler
cooling~\cite{Dalibard:1989p1149,cool:nobel98} that will become important
as the velocity of the atoms increases.

\subsection{Classical Approach}

In order to check the generality of the system and relevance of an
analysis in terms of a Fokker-Planck equation, as in
Eq.~(\ref{langevin}), we also perform a simple, completely classical
Monte-Carlo simulation for the Langevin equation corresponding to
Eq.~(\ref{langevin}), for a 1D system with classical particles in a
tilted washboard potential, as in Eq.~(\ref{washboard}). In this
simulation, we let the noise term and the friction scale linearly with
the potential depths. For simplicity and generalization we take a
uniform friction. The results are presented together with the
experimental data in Figs.~\ref{Fig_2} and \ref{Fig_3}. The basic
characteristics of our system are reproduced.

\section{Conclusion}

To conclude, we have made a quantitative study of how random isotropic
fluctuations, together with a very small bias force, gives rise to an
average drift. When the fraction between bias force and trapping
potential, $F/V_0$ is varied, the magnitude of this drift can
change. The system can be well described by a Fokker-Planck equation
formalism, and the experimental control, together with the precision
in the measurements, make the system suitable as a general testbed for
studies of fundamental fluctuation phenomena. To emphasize this, we
qualitatively reproduce our data with a simplified classical
simulation, as well as with a careful semi-classical Monte-Carlo
simulation of the laser cooling setup.  Our results also evidence the
`stop-and-go' nature of the dynamics of the atoms, where they
continuously exchange between being trapped in potential wells and
travelling over many
wells~\cite{Marksteiner:1996p1152,Ellmann:2001p1222,Dion:2005p1121,Jonsell:2006p1132}.

One of the constraints of the present experiment is the use of gravity
as the bias force.  This leaves the potential depth $V_0$ as the main
variable parameter, but it is only accessible through the laser
irradiance, which also modifies diffusion and friction.  One possible
solution would be the introduction of an additional laser beam, using
its radiation pressure as the bias force.  Moreover, it is possible to
operate the optical lattice far-detuned from the atomic resonance,
where it only serves as a (conservative) potential. Extra laser
fields could provide diffusion and friction, allowing the exploration
of a broad range of scenarios.

\acknowledgments

We thank S. Jonsell for stimulating discussions.  This project has
been supported by the Swedish Research Council, Knut \& Alice
Wallenbergs stiftelse and Carl Trygger stiftelse. The semi-classical
Monte-Carlo simulations were run at the National Supercomputing Center
(Link\"{o}ping).


\begin{thebibliography}{10}
\bibitem{gravity}
Note that we had previously mentionned in~\cite{Sjolund:2007p1141} that the
  effect of gravity was three orders of magnitude smaller \emph{than the recoil
  energy}, which is erroneous.
\end{thebibliography}

\end{document}